\documentclass[12pt]{article}

\usepackage{newtxtext,newtxmath}

\usepackage{graphicx}

\usepackage[letterpaper,margin=1in]{geometry}

\renewenvironment{abstract}
	{\quotation}
	{\endquotation}

\date{}

\makeatletter
\renewcommand{\fnum@figure}{\textbf{Figure \thefigure}}
\renewcommand{\fnum@table}{\textbf{Table \thetable}}
\makeatother

\usepackage{scicite}

\usepackage{url}

\def\scititle{
	Geese achieve stationary takeoff via synergistic wing kinematics and enhanced aerodynamics
}
\title{\bfseries \boldmath \scititle}

\author{
	Jinpeng~Huang$^{1}\dagger$,
	Yang Xiang$^{1\ast}\dagger$,
	Lunbing Chen$^{1}$,
        Suyang Qin$^{1,2}$,\\
        Jixin Lu$^{1}$,
        Sen Ye$^{1}$,
        Yong Chen$^{1,3}$,
        Hong Liu$^{1\ast}$\and
	\small$^{1}$J.C. Wu Center for Aerodynamics, School of Aeronautics and Astronautics,\and \small  Shanghai Jiao Tong University, Shanghai \& 200240, China.\and
    \small $^{2}$ Bird Aerospace Technology (Suzhou) Co., Ltd., Suzhou \& 215412, China. \and
    \small $^{3}$ Commercial Aircraft Corporation of China, Shanghai \& 200126, China. \and
	\small$^\ast$Corresponding author. Email: xiangyang@sjtu.edu.cn; hongliu@sjtu.edu.cn\and
	\small$^\dagger$These authors contributed equally to this work.
}

\begin{document} 

\maketitle

\begin{abstract} \bfseries \boldmath
Stationary take-off, without a running start \cite{usherwood2003aerodynamics} or elevated descent \cite{kleinheerenbrink2022optimization}, requires substantial aerodynamic forces to overcome weight, particularly for large birds such as geese exceeding 2 kg. However, the complex wing motion and high-Reynolds-number (Re $\approx$$10^5$) flow dynamics challenge conventional expectations of avian flight aerodynamics, rendering this mechanism elusive. Analyzing 578 stationary take-offs from seven geese (\textit{Anser cygnoides}) and applying Principal Component Analysis (PCA), we reveal that the complex wing kinematics collapse onto a low-dimensional manifold dominated by two synergies: a Stroke Synergy responsible for fundamental rhythmic stroke, and a Morphing Synergy governing spanwise geometry. This modular control strategy orchestrates a stereotyped wing kinematics featuring an accelerated translational downstroke and a rapid tip-reversal upstroke. By integrating wing kinematic analysis with the mass distribution of the geese, we quantified the aerodynamic forces and found that entirely positive lift and thrust are generated throughout the motion cycle. The enhanced aerodynamic performance of geese takeoff results from three principal mechanisms. During the downstroke, significant lift generated from wing acceleration is predicted by the quasi‑steady framework. Flow visualization reveals that wake capture further enhances the lift generation in downstroke by orienting the position of wake vortices. During the upstroke, the distal wing performs a rapid pitching motion and generates a substantial thrust, the vertical component of which contributes significantly to weight support.
\end{abstract}

\begin{figure*}[t]
	\centering
	\includegraphics[width=12cm]{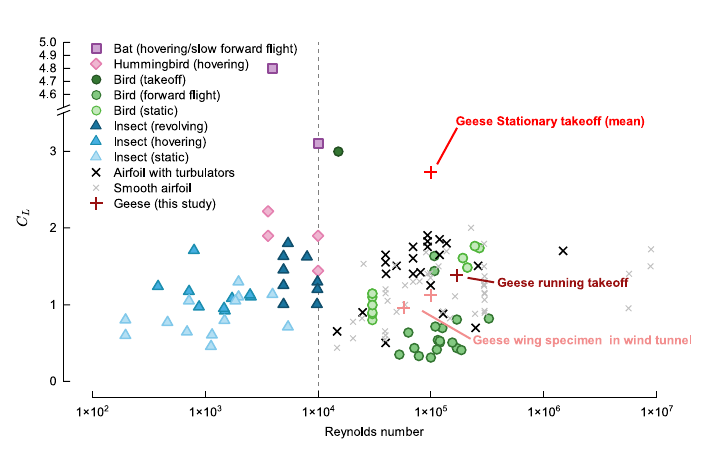} 
    \caption{\textbf{Aerodynamic regimes and the high-lift paradox of goose stationary takeoff.}  
    An overview of maximum lift coefficients ($C_{L,\max}$) and lift coefficients ($C_{L}$) under typical flight conditions reported in the literature across a range of Reynolds numbers ($Re$). Here, the suffix 'static' (across birds, bats, and insects) denotes data obtained under static flow conditions. For insects, 'revolving wings' refers to propeller-like experiments. Across all species, 'forward flight' and 'hovering' represent motion data derived from in vivo observations or motion capture systems. Notably, the data points of geese stationary takeoff are derived from typical morphological ranges ($m=2.1\text{--}3.5 \text{ kg}$, $S=0.296\text{--}0.36 \text{ m}^2$) and a characteristic wing velocity of $U \approx 7 \text{ m/s}$. The estimates yield a Reynolds number of $Re \approx 9.6 \times 10^4$ and a required lift coefficient range of $C_L \approx 1.9\text{--}3.9$. Detailed calculation methods and data sources are provided in the Supplementary Information.
    }
	\label{fig:0} 
\end{figure*}

\noindent
Flight has evolved independently across diverse lineages, yet the physical imperative remains a universal law: obtaining sufficient aerodynamic forces to overcome gravity and aero drag. 
This law is typically framed within the constraints of Reynolds number ($Re$) scaling (Fig. \ref{fig:0}). 
At low Reynolds numbers ($Re \sim 10^2-10^4$), small fliers (insects, hummingbirds and small bats) exploit unsteady mechanisms, such as leading-edge vortices and wake capture, to achieve high lift coefficients ($C_{L,\max} > 2$), to address the inadequacy of steady aerodynamics for producing sufficient lift\cite{2002Unconventional,ellington1996leading, lentink2009rotational, chin2016flapping, eldredge2019leading, jones2010unsteady, song2014three, altshuler2004aerodynamic, achache2017hovering, bullenquasi, norberg1976aerodynamics}. 
Conversely, at higher Reynolds numbers ($Re > 10^4$), characteristic of larger birds and conventional aircraft, flight is generally governed by steady aerodynamics, and increasing lift primarily involves increasing speed and the lift coefficient \cite{pennycuick2001speeds, withers1981aerodynamic,omar2020numerical,usherwood2010aerodynamic,usherwood2003aerodynamics,eldredge2019leading,jones2010unsteady}. 


However, stationary takeoff, an extreme survival capability allowing ground-to-aerial transition  without external acceleration from running or gravity, presents a striking aerodynamic paradox in species such as geese, apparently defying this conventional dichotomy. Figure \ref{fig:0} shows geese achieve stationary take-off operating at high Reynolds numbers ($Re \approx 10^5$) and generating exceptionally high lift coefficients ($C_{L,\text{mean}} > 2.5$), substantially higher than the maximum values obtained from wing-specimen wind-tunnel tests ($C_{L,\max} = 1.2$, figure S1). This suggests that large birds, during the critical phase of stationary takeoff, must exploit extra high-lift mechanisms at scales previously thought to be dominated by steady flows. 
By contrast, the running takeoff of geese requires a lift coefficient ($C_{L} = 1.38$) aligns with the steady aerodynamics, where lift increases with forward velocity \cite{usherwood2003aerodynamics}. 

In addition to the aerodynamic challenges, the kinematics of the wing during a stationary takeoff are also cognitively challenging. Existing studies on avian take-off have largely focused on gross parameters such as stroke plane angles and wingtip trajectories \cite{berg2010wing, chin2019birds, 1996Tobalske}. 
However, unlike the relatively rigid wings of insects or hummingbirds, which can be approximated as flat plates \cite{dickinson1999wing}, the avian wing is a multi-hinged, highly flexible structure capable of complex morphing. Crucially, the functional link between this intricate wing kinematic and the generation of aerodynamic forces beyond steady limits during the critical, low-speed stationary takeoff remains obscure.

\begin{figure*}[t]
	\centering
	\includegraphics[width=17.8cm]{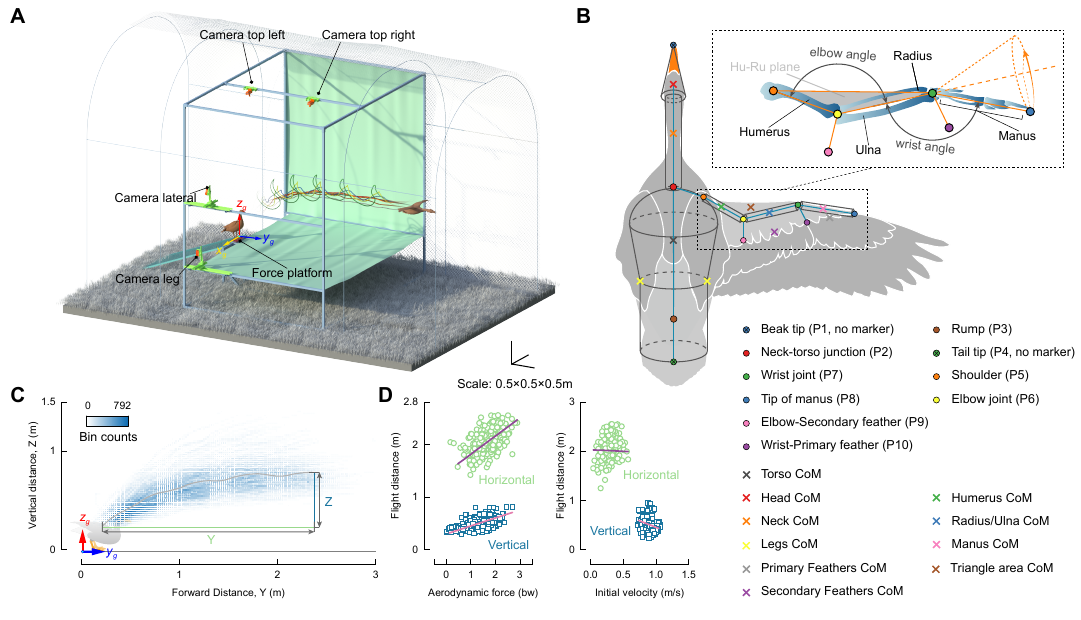} 
    \caption{\textbf{Motion capture and anatomical reconstruction of goose take-off.}  
    (\textbf{A}) Experimental setup for synchronized force and motion capture. Geese were trained to initiate take-off from a force platform within a calibrated motion capture system. 
    (\textbf{B}) Wing kinematics were driven by the humerus, radius/ulna, and manus, reconstructed from 10 anatomical key points on both torso and wings (8 of which were marked; circular dots). The goose was modeled as a composite geometric system, allowing real-time calculation of the CoM (denoted by cross) based on key point positions and anatomical measurements.
    (\textbf{C}) Spatial distribution of CoM trajectories across all $n=578$ take-off trials ($n=7$ geese). Bin counts represent the number of CoM samples falling within a 2 cm × 2 cm space in the lateral view. 
    (\textbf{D}) Horizontal (green cycle) and vertical (blue square) displacements of the CoM plotted against initial lift-off velocity and mean aerodynamic force within three wingbeat cycles. Solid lines show linear regressions in the vertical (pink) and horizontal (purple) directions, respectively.}
	\label{fig:1} 
\end{figure*}

Here, to investigate stationary take-off in geese (\textit{Anser cygnoides}), a cohort of \(n=36\) individuals from a farm in Suzhou, eastern China, were trained over 6 mouths to take off from a standstill. We selected $n=7$ well-trained individuals ($3.17 \pm 0.30$~kg) for this study (table S1, see Bird and Training). The custom-built outdoor experimental setup (Fig.\ref{fig:1}A, movie S2) included an instrumented hindlimb force platform (HFP) to measure hindlimb-generated forces (see Hindlimb Force Platform) and a 3D motion capture system (MCS) to track the 3D positions of 10 key anatomical points over 4–5 wingbeats post-lift-off (see Motion Capture System). Four markers were placed on the torso and six on wings(Fig. \ref{fig:1}B). This setup enabled detailed reconstruction of wing kinematics, including musculoskeletal movements and wing surface deformations (Fig.\ref{fig:2}B, movie S3, see Wing Kinematic). Another similar take-off setup suitable for flow visualization was constructed indoors to capture both local and global flow features during goose take-off (see Flow Pattern Visualization). Additionally, a mass distribution model (MDM, see Mass Distribution Model) was developed using mass and morphological data (see Supplementary Information). This model approximated the goose as a system of basic geometric objects, allowing determination of the center of mass (CoM) and mass of individual body segments (Fig.\ref{fig:1}B). Integrating MDM with acceleration data from kinematic differentiation provided further insights into take-off dynamics. Finally, kinematic and dynamic results were validated through bone length measurements and a customized acceleration calibration plate (see Aerodynamic Force and Validation). 

\section*{Results}

Data from \(n=578\) take-off flights performed by \(n=7\) geese were collected. The torso CoM for each goose was determined using the MDM, and lateral spatial histograms of individual and aggregated CoM trajectories were constructed to visualize takeoff trajectories (Fig.~\ref{fig:1}C). The distribution of post lift-off trajectories was broad, with individual variation, indicating the distinct take-off choices adopted by geese (figure S2). Stationary take-off begins with a transition from standing to crouching (movie S1), which is defined as the moment when the vertical ground reaction force of the hindlimbs falls to 95\% of body weight (bw), followed by leg extension and wing unfolding to launch upward and forward. First downstroke/upstroke transition occurs within $39.82\pm39.85$~ms after lift-off, we excluded the brief interval and considered the transition point itself as the moment of lift-off. The torso pitch angle stabilizes at $21.22 \pm 0.50^\circ$ after the third wingbeat (figure S3), which we define as the end of the stationary takeoff.

\subsection*{Stereotyped kinematics driven by two synergies}

Post–lift-off flight trajectories exhibit significant variations (Fig.~\ref{fig:1}C), which are found to be predominantly determined by hindlimb-generated initial velocity rather than aerodynamic modulation ( see Factors Affecting the Take-off Trajectory). Flight distances over the first three wingbeats correlate strongly with initial velocity in both horizontal ($r_H=0.649$) and vertical ($r_V=0.739$) directions, whereas correlations with the mean aerodynamic acceleration are negligible ($r_H=-0.025$; $r_V=-0.229$). These results suggest that the wings function as a stable aero-force generator, necessitating a highly stereotyped kinematic output immediately upon lift-off.

\begin{figure*}[t]
	\centering
	\includegraphics[width=17.8cm]{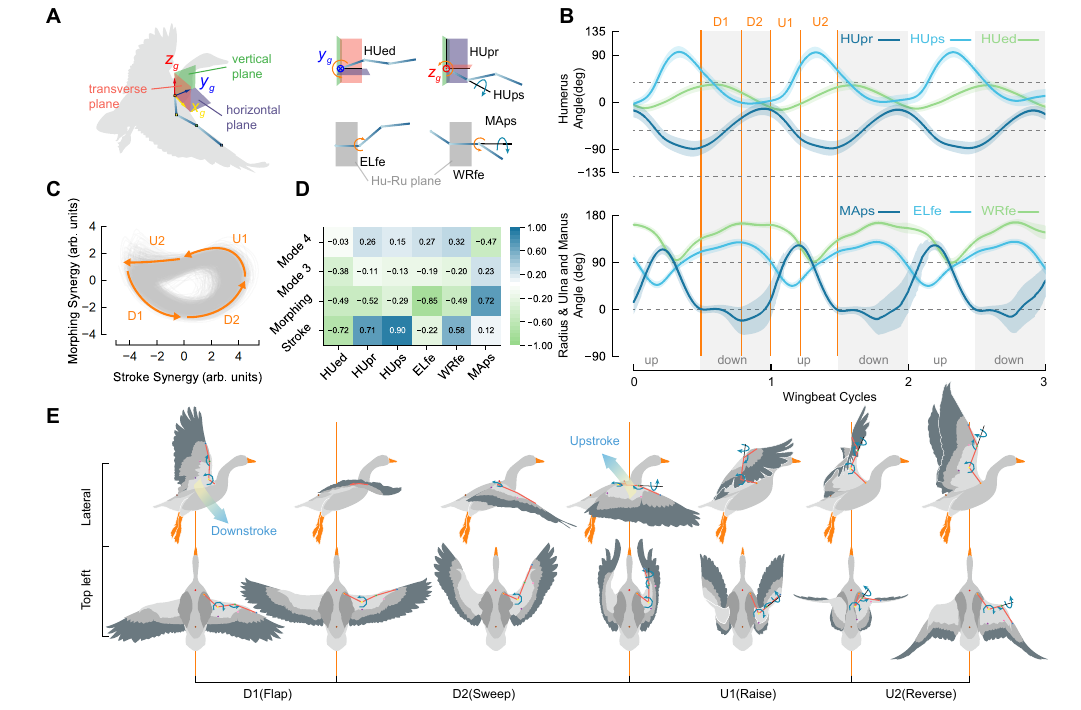} 
	\caption{\textbf{Stereotyped take-off wing kinematics.}  
    (\textbf{A}) Schematic diagram of the degrees of freedom of the goose wing. Wing motion was described relative to the shoulder (P5) in global reference frame. The humerus exhibited three DoFs: elevation/depression (HUed), pronation/supination (HUps), and protraction/retraction (HUpr). Elbow flexion/extension (ELfe) was enabled by the radius/ulna, while wrist flexion/extension (WRfe) and pronation/supination (WRps) were facilitated by the manus. 
    (\textbf{B}) Time history of wing skeletal motion. 
    (\textbf{C}) Superimposed phase space trajectories of the Stroke Synergy versus the Morphing Synergy across all experimental trials. 
    (\textbf{D}) Correlation matrix between the PCA modes derived from wing marker coordinates and the six anatomical DoFs.
    (\textbf{E}) Lateral and top views of a complete wingbeat. A typical wingbeat cycle from downstroke to upstroke consists of four key phases—\textit{flap}, \textit{sweep}, \textit{raise}, and \textit{reverse}—are marked in (B, C and E), with the orange line marks the boundaries of the each phases. The gray area marks the downstroke.}  
	\label{fig:2} 
\end{figure*}

To elucidate the coordination underlying the stereotyped wing motion, we performed Principal Component Analysis (PCA) on the wing markers in the body frame. The analysis reveals that wing kinematics collapse onto a low-dimensional manifold dominated by two synergies (together explaining 93.6\% of the total variance). Wing kinematics are characterized by six angles describing the degrees of freedom (DoFs) of the humerus, ulna/radius, and manus relative to the shoulder (Fig. \ref{fig:2}A-B, see Wing Kinematics). To physically understand the identities of the PCA-derived synergies, we analyzed their correlations with these anatomical DoFs (Fig. \ref{fig:2}D). The first synergy identifies as a Stroke Synergy, defined by a rigid coupling of humeral depression, protraction, and pronation ($r_{\text{HUed}} \approx -0.72, r_{\text{HUpr}} \approx 0.71, r_{\text{HUps}} \approx 0.90$). The second synergy functions as a Morphing Synergy, governed primarily by coordinated flexion ($r_{\text{ELfe}} \approx -0.85, r_{\text{WRfe}} \approx -0.49$) and manus pronation ($r_{\text{MAps}} \approx 0.72$) to regulate spanwise geometry. 

A typical wingbeat cycle is defined by the temporal superposition of these two synergies (Fig. \ref{fig:2}C-D). Stroke Synergy serves as the rhythmic driver, oscillating with high amplitude throughout both downstroke and upstroke to dictate the global humeral motion, peaking at the stroke reversals. Superimposed on Stroke Synergy is the modulation of Morphing Synergy, which regulates wing morphing and displays distinct characteristics across the wingbeat phases. Based on the two synergies, together with the vertical position of the wingtip marker (P8), the cycle is divided into four phases: D1 (\textit{flap}) and D2 (\textit{sweep}) during the downstroke, followed by U1 (\textit{raise}) and U2 (\textit{reverse}) during the upstroke (Fig.~\ref{fig:2}B, C and E). The peaks of vertical position of the wingtip marker (P8) define the stroke reversal, and the zero-crossings of the Stroke Synergy define the sub-phase transitions, i.e., D1-D2 and U1-U2. This synergy-based segmentation is physically robust, as it coincides with the sharp turning points in the joint angle trajectories, marking distinct shifts in the underlying kinematic strategy (see Wing Kinematic). During the downstroke, Stroke Synergy transitions from negative to positive, driving the wing protraction (HUpr increasing) and depression (HUed decreasing) and protraction (HUps decreasing). Concurrently, Morphing Synergy exhibits a distinct trajectory reversal. It decreases rapidly in the \textit{flap} phase (D1) to drive maximal wing extension (ELfe increasing to $128.4\pm6.3^\circ$; WRfe increasing to $164\pm4.2^\circ$), then rises in the \textit{sweep} phase (D2) to initiate distal flexion (ELfe decreasing to $91.3\pm10.6^\circ$; WRfe constant) and manus supination (MAps increasing to $21.4\pm29.3^\circ$). During the upstroke, Stroke Synergy transitions from positive to negative, driving global elevation (HUed increasing), retraction (HUpr decreasing), and supination (HUps increasing). Concurrently, Morphing Synergy continues to increase, driving the wing into a deeper fold. It rises in the \textit{raise} phase (U1) to create a compact fold (ELfe decreasing to $51.8\pm7.4^\circ$; WRfe to $126.5\pm9^\circ$; MAps reaching $123.2\pm8^\circ$), then maintains a high plateau in the \textit{reverse} phase (U2) to lock the folded state for the wingtip reversal maneuver.

\subsection*{Regulation of aerodynamic and inertial forces}

\begin{figure*}[t]
    \centering
    \includegraphics[width=17.8cm]{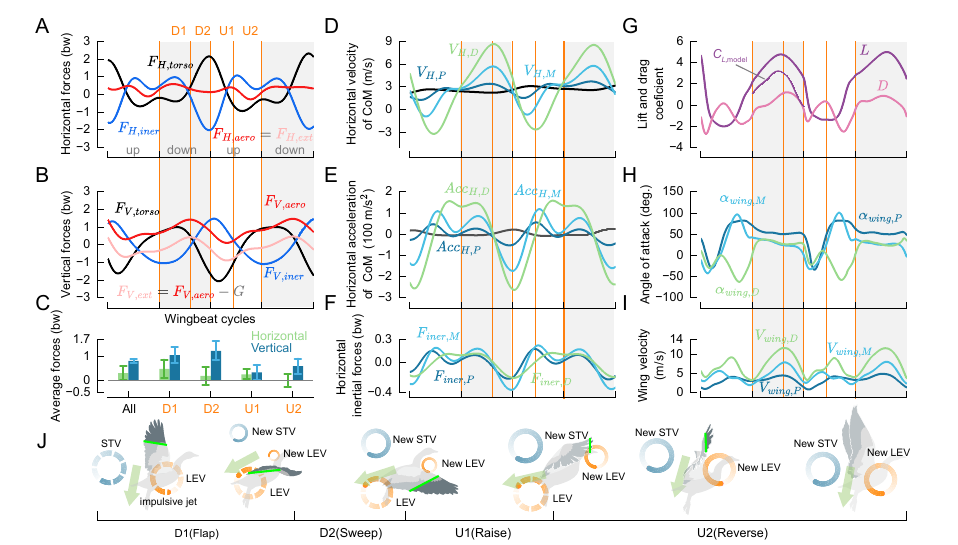}
    \caption{\textbf{Aerodynamic force generation and wing kinematics.} 
    \textbf{A, B,} Time-resolved aerodynamic forces ($F_{\text{aero}}$), wing inertial forces ($F_{\text{iner}}$), and torso inertial forces ($F_{\text{torso}}$) in the horizontal (\textbf{A}) and vertical (\textbf{B}) directions. Specific aerodynamic forces (red lines) were resolved by analytically decoupling wing inertial contributions (blue lines) from the total translational kinetics (black lines). 
    \textbf{C,} Mean aerodynamic forces in horizontal (light blue) and vertical (dark blue) directions across four phases (flap, sweep, raise, reverse) and the full wingbeat cycle. 
    \textbf{D--F,} Horizontal kinematics and kinetics of wing segments relative to the torso: relative velocity (\textbf{D}), relative acceleration (\textbf{E}), and derived inertial forces based on segmental mass (\textbf{F}). Data are presented for three representative segments: humerus (proximal, P), radius-ulna (middle, M), and manus (distal, D); inertial contributions from other wing elements were negligible. 
    \textbf{G,} Decomposition of instantaneous aerodynamic forces based on wing velocity at the radius of gyration. Solid lines represent measured lift (purple) and drag (pink); The thin purple line indicate theoretical predictions from quasi-steady models. 
    \textbf{H, I,} Angle of attack (AoA) (\textbf{H}) and velocity (\textbf{I}) of the proximal, middle, and distal wing sections. 
    \textbf{J} Schematic of flow structures observed from flow visualization, showing the generation of a starting vortex (STV, blue loop) and a leading-edge vortex (LEV, orange loop) during the wingbeat. Solid loops represent vortices from the current wingbeat cycle, while dashed loops indicate those from the previous cycle. The schematic is based on experimental flow visualization results (see Flow Pattern Visualization). 
    Gray shading indicates the downstroke; orange vertical lines delimit the four wingbeat phases.}
    \label{fig:4}
\end{figure*}

 
Using periodic wing kinematics, geese dynamically regulate both inertial and aerodynamic forces of their massive wings (\(14.12 \pm 1.38\%\) of body weight; figure~S4) during stationary take-off. As the wings accelerate downward and forward, the torso experiences acceleration (\(a_{\text{torso}}\)) from the net external force (\(F_{\text{ext}} = F_{\text{aero}} + G\)) on the whole-body CoM, while the wings generate additional upward-backward inertial forces (\(F_{\text{iner}}\)) due to their motions relative to the torso. These wing inertial and torso forces (\(F_{\text{torso}} = ma_{\text{torso}}\)) allow estimating the variations of aerodynamic forces (Fig.~\ref{fig:4}A, B)~(see Aerodynamic Force and Validation).

Across three wingbeat cycles post lift-off, average vertical aerodynamic force is \(\bar{F}_{\text{V}} = 0.81\pm0.09\)~bw, and average horizontal aerodynamic force is \(\bar{F}_{\text{H}} = 0.30\pm0.28\)~bw (Fig.~\ref{fig:4}C), accounting for the observed decrease in vertical velocity (42.6\%) and increase in horizontal velocity (90.6\% ). 
All four phases contribute positive vertical force, with downstroke dominating (\(\bar{F}_{\text{V}}^\text{D1} = 1.05\pm0.31\)~bw; \(\bar{F}_{\text{V}}^\text{D2} = 1.22\pm0.39\)~bw). 
Moreover, \textit{flap}, \textit{sweep} and \textit{raise} phases also generate positive horizontal aerodynamic forces with averages of \(\bar{F}_{\text{H}}^\text{D1} = 0.48\pm0.38\)~bw,  \(\bar{F}_{\text{H}}^\text{D2} = 0.20\pm0.37\)~bw, and \(\bar{F}_{\text{H}}^\text{U1} = 0.25\pm0.21\)~bw, with two peaks shown in \textit{flap} and \textit{raise} phases (Fig.~\ref{fig:4} A). The \textit{reverse} phase sometimes also yields positive thrust (\(\bar{F}_{\text{H}}^\text{U2} = -0.01\pm0.26\)~bw). 

Stroke Synergy generates periodic inertial force oscillations in both horizontal and vertical directions (Fig. \ref{fig:4}A, B), with the cycle-averaged inertial force approaching zero ($|\bar{F}_{\text{iner}}| < 0.01~\text{bw}$). Morphing Synergy exhibits a distinct upstroke-downstroke asymmetry that mitigates adverse intra-cycle inertial effects, particularly for the horizontal component of the upstroke-to-downstroke transition (D2U1 transition), where peak inertial force is reduced by $54\% \pm 13\%$ relative to the downstroke-to-upstroke transition (U2D1 transition, Fig. \ref{fig:4}A). This mitigation is achieved via a proximal-to-distal acceleration wave during the reversal phase (U2) that enables proximal segments to utilize a minor acceleration to rapidly synchronize velocity with the torso, whereas distal segments exhibit sustained acceleration to execute a backward flick (Fig. \ref{fig:4}D and E). Crucially, despite this high distal acceleration, the system adheres to a mass-dominated inertial regime where the heavier proximal segments dictate the force profile (Fig. \ref{fig:4}F). This generates a primary inertial peak ($F_{\text{iner}}=0.93\pm 0.27~\text{bw}$) followed by a distinct valley ($F_{\text{iner}}=0.30\pm 0.28~\text{bw}$) (Fig. \ref{fig:4}A). As the reversal concludes and the downstroke initiates, the wing accelerates forward, producing a second positive peak ($F_{\text{iner}}=0.98\pm 0.17~\text{bw}$) (Fig. \ref{fig:4}A).

This kinematic staggering, where proximal braking precedes distal turnover, sculpts the horizontal inertial force during upstroke-to-downstroke transition into a characteristic ‘saddle’ profile. Functionally, this mechanism distributes the impulsive wing inertial load over time, effectively shaving off the singular high-magnitude spike inherent to rigid reversals to minimize the instantaneous inertial impact on the torso.

\subsection*{Aerodynamic force generation mechanisms}

To uncover generation mechanisms of aerodynamics forces, we decomposed the measured instantaneous forces relative to the wing's instantaneous velocity, thereby obtaining the lift and drag, and analyzed them in conjunction with both a quasi-steady model and flow visualizations (Fig. \ref{fig:4}G and J; see Aerodynamic Force and Validation). Lift and drag are perpendicular to and opposite to the wing's instantaneous velocity, respectively. 

During the downstroke, the peak lift and drag appear at the D1-D2 transition (Fig. \ref{fig:4}G–H; \( C_{L}^{\text{max}} = 4.79 \pm 1.45\), \(C_{D}^{\text{max}} = 1.21 \pm 1.63\)), where the Morphing Synergy is at its minimum, representing a state of maximum wing extension and unfolding. At this moment, lift acts upward and forward, while drag acts upward and backward, meaning both possess vertical components contributing to weight support (figure S5). Lift provides \(\sim89\%\) of vertical support and \(\sim200\%\) of net forward force; drag adds \(\sim11\%\) to vertical support but fully offsets forward motion with \(\sim100\%\) backward force. During the wing flaps, angles of attack (AoAs) remain nearly stable across all sections (Fig. \ref{fig:4}H; \(\bar{\alpha}_{\text{P}} = 54.1 \pm 4.5^\circ\), \(\bar{\alpha}_{\text{M}} = 28.3 \pm 9.2^\circ\), \(\bar{\alpha}_{\text{D}} = 29.9 \pm 5.8^\circ\)), while the velocities of proximal, middle, and distal wings (defined as the velocity at the CoM of wing sections) rise then fall(Fig. \ref{fig:4}I), matching aerodynamic force trends. Thereby, it suggests that the aerodynamic force should be generated by the accelerating starting motion of wing at high AoA. 
To further quantify this, we applied a theoretical model based on starting-plate that decomposes the instantaneous lift coefficient into non-circulatory ($C_{L_{\text{NC}}}$, added-mass effects) and circulatory ($C_{L_{\Gamma}}$, vortex growth) components (see Theoretical Aerodynamic Force Model):
\begin{equation}
C_{L,\text{model}}(t) = C_{L_{\text{NC}}}(t) + C_{L_{\Gamma}}(t).
\end{equation}
This model predicts a progressive rise in lift from $C_{L,\text{model}} \approx 1.11$ at the onset of the \textit{flap} phase to a peak of $C_{L,\text{model}} = 3.24$ near the transition, followed by a decline to zero (thin purple line in Fig. \ref{fig:4}G). However, these values remain substantially lower than the measured results ($C_{L,\text{max}} \approx 4.79$) over the same interval. Since the model accounts for the intrinsic lift generation of an accelerating wing (including the formation of the starting vortex), the remaining discrepancy points to extrinsic unsteady mechanisms. 

The discrepancy between theoretical model $C_{L,\text{model}}$ and measured result $C_L$ is observed throughout the entire downstroke. Flow visualization gives the evolution of flow patterns (Fig. \ref{fig:4}J). Integrating the flow patterns, we attribute this gap primarily to wake capture, where the wing encounters and interacts with the residual LEV from the previous cycle, the wake-capture effect not captured by the starting-plate model \cite{2014Mechanism,2018On}. During downstroke, high AOA accelerating starting motion of wing generates a new LEV attached on the wing surface. Wing morphing in upstroke, particular the wingtip reversal behavior, reorients the position of the new LEV, thereby providing the opportunity for wake capture in the next wingbeat cycle. 

During the upstroke, two key features are observed. First, owing to the three wing sections flip ventral-side up (wing plane angles exceed $90^\circ$), lift becomes negative (toward the ventral side) at mid-upstroke (Fig. \ref{fig:4}G-H), yet still contributes \(\sim154\%\) to vertical support and \(\sim85\%\) to net forward force(figure~S5). Second, drag reverses the direction during the U1 and U2 phases (Fig. \ref{fig:4}G-H), with two negative peaks (\(C_{D}^{\text{first}} = -1.54\pm1.28\), \(C_{D}^{\text{second}} = -2.5\pm1.8\)), supporting vertical force (\(\sim80\%\) and \(\sim87\%\)) and providing all net forward force in these moments (figure S5).
Examining the variations of AoAs, we observe a functional partitioning across the wing. The proximal and middle sections undergo a drastic AoA excursion, shifting from negative ($\sim-30^{\circ}$) to positive ($\sim80^{\circ}$), a kinematic signature consistent with a deep dynamic stall process \cite{2015Dynamic}. This stall state facilitates the formation of a strong LEV , thereby augmenting instantaneous lift (Fig. \ref{fig:4}J). In contrast, the distal wing functions primarily as a propulsion generator, executing fast pitching oscillations ($\sim18 \, \text{Hz}$) with amplitude increasing from $33^\circ$ to $69^\circ$. To determine if this specific motion accounts for the observed reversal of drag into thrust, we applied the high-Reynolds-number scaling law for oscillating airfoils to quantify the resulting mean thrust ($\bar{C}_T$) \cite{floryan2017scaling,dave2020variable,green2008effects}:
\begin{equation}
\bar{C}_T \approx \beta St^2 - C_D.
\end{equation}
where $C_D$ is the quasi-steady offset drag term, $St$ is the Strouhal number and $\beta$ is an empirical coefficient (see Theoretical Aerodynamic Force Model). The model yields a theoretical thrust coefficient of $\bar{C}_T=0.87$, which aligns remarkably well with the experimental measurement ($\bar{C}_D=-0.96$, noting $C_T = -C_D$). This quantitative agreement corroborates that the propulsive force is mechanistically driven by the rapid pitching dynamics of the distal wing.

\section*{Conclusion}
Geese achieve stationary take-off via a low-dimensional strategy that reduces complex kinematics into two fundamental synergies: a Stroke Synergy driving the rhythmic beat, and a Morphing Synergy modulating wing shape. The dual-synergy kinematics of the wings allows for a quantifiable division of the wingbeat cycle into four distinct phases: flap, sweep, raise, and reverse. This kinematic pattern generates positive vertical lift throughout all four phases to overcome gravity, while simultaneously, wing morphing substantially dampens the peak horizontal inertial forces.  

Force analysis suggests that the enhanced aerodynamics is mainly attributed to three mechanisms: high AOA accelerating, wake capture, and rapid pitching. During downstroke, the high AOA accelerating motion of geese's wings generates a high lift coefficient, which is predicted by the theoretical quasi-steady model. The accelerating force is not sufficient alone to explain the measured high force during downstroke. Wake capture is observed from flow visualization to account for the gap between the theoretical forces and the measured forces. The finding of wake capture in such high Re number ($>10^5$) flight of geese is quite surprise, and the manner of wake capture by wing morphing is also novel.  During the upstroke, the distal wing performs a rapid pitching motion, which converts drag into thrust, further providing the vertical force to support weight. 

Finally, our results show that stationary take-off in large birds emerges from a conserved, synergy-based kinematic framework that robustly orchestrates aerodynamic enhancement. This paper not only elucidates the dynamics of stationary takeoff in large avian species, but also provides actionable kinematic and aerodynamic control strategies for designing bio-inspired aerial systems capable of takeoff regime transition. 


\clearpage 

%
\bibliography{refs} 

\begin{thebibliography}{10}
\providecommand{\url}[1]{\texttt{#1}}
\expandafter\ifx\csname urlstyle\endcsname\relax
  \providecommand{\doi}[1]{doi:\discretionary{}{}{}#1}\else
  \providecommand{\doi}{doi:\discretionary{}{}{}\begingroup \urlstyle{rm}\Url}\fi

\bibitem{usherwood2003aerodynamics}
J.~R. Usherwood, T.~L. Hedrick, A.~A. Biewener, The aerodynamics of avian take-off from direct pressure measurements in Canada geese (Branta canadensis). \emph{Journal of Experimental Biology} \textbf{206}~(22), 4051--4056 (2003).

\bibitem{kleinheerenbrink2022optimization}
M.~KleinHeerenbrink, L.~A. France, C.~H. Brighton, G.~K. Taylor, Optimization of avian perching manoeuvres. \emph{Nature} \textbf{607}~(7917), 91--96 (2022).

\bibitem{2002Unconventional}
R.~B. Srygley, A.~L.~R. Thomas, Unconventional lift-generating mechanisms in free-flying butterflies. \emph{Nature} \textbf{420}~(6916), 660--664 (2002).

\bibitem{ellington1996leading}
C.~P. Ellington, C.~Van Den~Berg, A.~P. Willmott, A.~L. Thomas, Leading-edge vortices in insect flight. \emph{Nature} \textbf{384}~(6610), 626--630 (1996).

\bibitem{lentink2009rotational}
D.~Lentink, M.~H. Dickinson, Rotational accelerations stabilize leading edge vortices on revolving fly wings. \emph{Journal of experimental biology} \textbf{212}~(16), 2705--2719 (2009).

\bibitem{chin2016flapping}
D.~D. Chin, D.~Lentink, Flapping wing aerodynamics: from insects to vertebrates. \emph{Journal of Experimental Biology} \textbf{219}~(7), 920--932 (2016).

\bibitem{eldredge2019leading}
J.~D. Eldredge, A.~R. Jones, Leading-edge vortices: mechanics and modeling. \emph{Annual Review of Fluid Mechanics} \textbf{51}~(1), 75--104 (2019).

\bibitem{jones2010unsteady}
A.~Jones, H.~Babinsky, Unsteady lift generation on rotating wings at low Reynolds numbers. \emph{Journal of Aircraft} \textbf{47}~(3), 1013--1021 (2010).

\bibitem{song2014three}
J.~Song, H.~Luo, T.~L. Hedrick, Three-dimensional flow and lift characteristics of a hovering ruby-throated hummingbird. \emph{Journal of The Royal Society Interface} \textbf{11}~(98), 20140541 (2014).

\bibitem{altshuler2004aerodynamic}
D.~L. Altshuler, R.~Dudley, C.~P. Ellington, Aerodynamic forces of revolving hummingbird wings and wing models. \emph{Journal of zoology} \textbf{264}~(4), 327--332 (2004).

\bibitem{achache2017hovering}
Y.~Achache, N.~Sapir, Y.~Elimelech, Hovering hummingbird wing aerodynamics during the annual cycle. I. Complete wing. \emph{Royal Society Open Science} \textbf{4}~(8), 170183 (2017).

\bibitem{bullenquasi}
R.~Bullen, N.~McKenzie, Quasi-Steady Aerodynamic Model for Bat Airframes .

\bibitem{norberg1976aerodynamics}
U.~Norberg, Aerodynamics of hovering flight in the long-eared bat Plecotus auritus. \emph{Journal of Experimental Biology} \textbf{65}~(2), 459--470 (1976).

\bibitem{pennycuick2001speeds}
C.~Pennycuick, Speeds and wingbeat frequencies of migrating birds compared with calculated benchmarks. \emph{Journal of Experimental Biology} \textbf{204}~(19), 3283--3294 (2001).

\bibitem{withers1981aerodynamic}
P.~C. Withers, An aerodynamic analysis of bird wings as fixed aerofoils. \emph{Journal of Experimental Biology} \textbf{90}~(1), 143--162 (1981).

\bibitem{omar2020numerical}
A.~Omar, R.~Rahuma, A.~Emhemmed, Numerical Investigation on Aerodynamic Performance of Bird’s Airfoils. \emph{Journal of Aerospace Technology and Management} \textbf{12}, e4620 (2020).

\bibitem{usherwood2010aerodynamic}
J.~R. Usherwood, The aerodynamic forces and pressure distribution of a revolving pigeon wing, in \emph{Animal locomotion} (Springer), pp. 429--441 (2010).

\bibitem{berg2010wing}
A.~M. Berg, A.~A. Biewener, Wing and body kinematics of takeoff and landing flight in the pigeon (Columba livia). \emph{Journal of Experimental Biology} \textbf{213}~(10), 1651--1658 (2010).

\bibitem{chin2019birds}
D.~D. Chin, D.~Lentink, Birds repurpose the role of drag and lift to take off and land. \emph{Nature communications} \textbf{10}~(1), 5354 (2019).

\bibitem{1996Tobalske}
B.~W. Tobalske, K.~P. Dial, Flight kinematics of black-billed magpies and pigeons over a wide range of speeds. \emph{Journal of Experimental Biology} \textbf{199}~(Pt 2), 263--280 (1996).

\bibitem{dickinson1999wing}
M.~H. Dickinson, F.-O. Lehmann, S.~P. Sane, Wing rotation and the aerodynamic basis of insect flight. \emph{science} \textbf{284}~(5422), 1954--1960 (1999).

\bibitem{2014Mechanism}
Z.~Fu, S.~Qin, H.~Liu, Mechanism of transient force augmentation varying with two distinct timescales for interacting vortex rings. \emph{Physics of Fluids} \textbf{26}~(1), 626--630 (2014).

\bibitem{2018On}
S.~Qin, H.~Liu, Y.~Xiang, On the formation modes in vortex interaction for multiple co-axial co-rotating vortex rings. \emph{Physics of Fluids} \textbf{30}~(1), 011901 (2018).

\bibitem{2015Dynamic}
T.~C. Corke, F.~O. Thomas, Dynamic Stall in Pitching Airfoils: Aerodynamic Damping and Compressibility Effects. \emph{Annual Review of Fluid Mechanics} \textbf{47}~(1), págs. 479--505 (2015).

\bibitem{floryan2017scaling}
D.~Floryan, T.~Van~Buren, C.~W. Rowley, A.~J. Smits, Scaling the propulsive performance of heaving and pitching foils. \emph{Journal of Fluid Mechanics} \textbf{822}, 386--397 (2017).

\bibitem{dave2020variable}
M.~Dave, A.~Spaulding, J.~A. Franck, Variable thrust and high efficiency propulsion with oscillating foils at high Reynolds numbers. \emph{Ocean Engineering} \textbf{214}, 107833 (2020).

\bibitem{green2008effects}
M.~A. Green, A.~J. Smits, Effects of three-dimensionality on thrust production by a pitching panel. \emph{Journal of fluid mechanics} \textbf{615}, 211--220 (2008).

\end{thebibliography}
\bibliographystyle{sciencemag}

%
%
%
%
%
%


\section*{Acknowledgments}
We gratefully acknowledge Yucai Xu and Huiying Zhu for their invaluable assistance in the domestication and care of the geese. We also thank Haotian Hang for helpful discussions on the flow patterns, and Xiaobin Huang for his support in refining the manuscript. Special thanks go to the Science and Technology Bureau of Taicang, Suzhou, the Dianzhan Village Committee of Chengxiang Town and Bird Aerospace Technology (Suzhou) Co., Ltd., for their generous support in providing field sites and assistance with animal quarantine and welfare. This work was supported by NSFC (National Natural Science Foundation of China) Projects (Grant Nos. 12202273)
\paragraph*{Funding:}
This work was supported by NSFC (National Natural Science Foundation of China) Projects (Grant Nos. 12202273)
\paragraph*{Author contributions:}
Conceptualization: J.H., Y.X., L.C., S.Q., S.Y., Y.C., and H.L. Data curation: J.H. Formal analysis: J.H. and J.L. Funding acquisition: Y.X., S.Q., H. L. Investigation: J.H. and Y.X. Methodology: J.H. Y.X. and L.C. Project administration: J.H. and Y.X.  Resources: J.H., Y.X., S.Q. and S.Y. Software: J.H. Supervision: Y.X., H.L. Validation: J.H. Y.X. and L.C. Visualization: J.H. Writing – original draft: J.H. Writing – review \& editing J.H. Y.X. and L.C.
\paragraph*{Competing interests:}
The authors declare that they have no competing interests.
\paragraph*{Data and materials availability:}
All data needed to evaluate the conclusions in the paper are present in the paper and/or the Supplementary Materials. Additional data related to this paper may be requested from the authors.


\subsection*{Supplementary materials}
Materials and Methods\\
Supplementary Text\\
Figs. S1 to S12\\
Tables S1 to S2\\
References \textit{(29-\arabic{enumiv})}\\ 
Movie S1 to S4\\
Data S1 to S4


\end{document}